\pdfoutput=1

\documentclass[11pt]{article}

\usepackage[final]{acl}

\usepackage{times}
\usepackage{latexsym}

\usepackage[T1]{fontenc}

\usepackage[utf8]{inputenc}

\usepackage{microtype}

\usepackage{inconsolata}

\usepackage{graphicx}

\usepackage{amsmath}
\usepackage[ruled,vlined]{algorithm2e}
\usepackage{algorithm2e}
\usepackage{float}
\usepackage{mdframed}
\usepackage{geometry} \usepackage{array, makecell}%
\setcellgapes{3pt}
\usepackage{bbm}

\title{PromptMind Team at EHRSQL-2024: Improving Reliability of SQL Generation using Ensemble LLMs}

\author{Satya K Gundabathula \\
  \texttt{satyakesav123@gmail.com} \\\And
  Sriram R Kolar \\
  \texttt{sriramrakshithkolar@gmail.com} \\}

\begin{document}
\maketitle
\begin{abstract}
This paper presents our approach to the EHRSQL-2024 shared task, which aims to develop a reliable Text-to-SQL system for electronic health records. We propose two approaches that leverage large language models (LLMs) for prompting and fine-tuning to generate EHRSQL queries. In both techniques, we concentrate on bridging the gap between the real-world knowledge on which LLMs are trained and the domain-specific knowledge required for the task. The paper provides the results of each approach individually, demonstrating that they achieve high execution accuracy. Additionally, we show that an ensemble approach further enhances generation reliability by reducing errors. This approach secured us 2nd place in the shared task competition. The methodologies outlined in this paper are designed to be transferable to domain-specific Text-to-SQL problems that emphasize both accuracy and reliability.
\end{abstract}

\section{Introduction}

Text-to-SQL technology translates natural language questions into executable SQL queries that can answer the questions using a provided database. A robust Text-to-SQL system could significantly increase productivity for anyone using databases by providing an easy-to-use natural language interface and reducing the need for expertise in different SQL dialects. These systems are particularly more valuable in domains where SQL knowledge is not essential, such as healthcare, where healthcare professionals like doctors, nurses, and hospital administrators spend a significant amount of time interacting with patient health records stored in databases.

In the era of Large Language Models (LLMs), the field of Text-to-SQL is gaining prominence as these models demonstrate impressive text generation capabilities without the need for fine-tuning. Introduced in 2017, WikiSQL \citep{zhongSeq2SQL2017} remains one of the largest datasets for Text-to-SQL and primarily caters to relatively simple queries. Subsequently, the SPIDER \citep{spider} and MULTI-SPIDER \citep{multispider} datasets were developed. These datasets posed challenges with complex queries that required an understanding of the database schema and support for various languages. BIRD-Bench was introduced to bridge the gap between research and real-world applications by providing large and imperfect databases \cite{bird}. These datasets are good representations of typical Text-to-SQL tasks. However, the healthcare domain differs from these generic datasets for the following reasons:
\begin{itemize}
    \item The questions asked by users maybe highly specialized and specific to the medical field.
    \item To answer such questions, systems must also possess an understanding of clinical terminology.
    \item Reliability is of paramount importance as errors can have serious consequences.
\end{itemize}

These differences present unique challenges for developing a reliable Text-to-SQL system for the healthcare domain. EHRSQL is the first dataset that closely captures the needs of hospital staff and serves appropriately for building and testing Text-to-SQL systems in the healthcare domain \cite{ehrsql}.

Our solution aims to create a Text to SQL system that emphasizes both reliability and accuracy. To achieve this, we divide the task into two phases:
\begin{itemize}
\setlength\itemsep{0em}
    \item SQL Generation
    \item SQL Validation
\end{itemize}

In the first stage, we focus on SQL generation employing different techniques that include prompting and fine-tuning of LLMs. In both approaches, we use the same prompting strategy to provide the LLM with database information and question-related context. Specifically, we use table schemas combined with sample column values as the database context, and similar questions from the training data as the task context. To identify similar questions from the training data, we employ an embedding-based similarity technique. Then, our goal is to maximize the LLM's ability to generate highly accurate SQL statements utilizing this approach.

There are several reasons why LLMs may fail to generate correct SQL for a given question. Some common reasons include:
\begin{itemize}
\setlength\itemsep{0em}
    \item Misinterpretation of question's intent
    \item Incorrect assumptions or hallucinations about the database's tables or columns
    \item Inaccuracies or hallucinations in the generated SQL query
\end{itemize}

Unlike many text generation tasks, Text-to-SQL tasks have a limited number of correct answers but potentially infinite incorrect ones. Inspired by this, we develop a second stage that evaluates the accuracy of the generated SQL. To evaluate the same, we propose an approach for Text-to-SQL that combines the results of multiple robust LLMs. Stronger LLMs often produce consistent outputs despite variations in temperature or other parameters, while smaller LLMs show lower consistency and accuracy. By leveraging the strengths of several robust LLMs, our approach minimizes the number of incorrect SQL queries and enhances the overall robustness and reliability of the Text-to-SQL system.

In the remainder of this paper, we discuss related work, introduce the EHRSQL-2024 task and dataset, and present our two-stage approach. We then provide the results of our experiments and conclude with a summary of our findings.

\section{Related Work}

Prior to the advent of LLMs, the primary focus of research in natural language processing involved refining specialized models using innovative strategies \cite{rat-sql}. Additionally, substantial efforts were devoted to developing sophisticated pre-training methodologies, such as those proposed by STAR \cite{star}, and exploring decoding strategies, as exemplified by PICARD \cite{picard}. However, these approaches typically require substantial computational resources and novel techniques.

Large Language Models (LLMs) have been trained extensively on textual data, which has equipped them with vast knowledge. As a result, they exhibit exceptional probabilistic reasoning abilities and can excel at various tasks even without explicit training. Zero-shot prompting techniques, when used with LLMs, have not only narrowed the performance gap on Text-to-SQL but have also surpassed specialized pre-trained or fine-tuned models. Several prompt techniques have been developed based on this zero-shot approach for Text-to-SQL tasks, leading to remarkable achievements on datasets such as SPIDER \cite{c3_0_shot}, \cite{0_shot}. Zero-shot generation capabilities can be further enhanced through techniques like in-context learning (ICL) and few-shot prompting.

DIN-SQL \cite{dinsql} adopts an in-context learning approach to break down complex SQL generation into manageable sub-tasks, leading to improved performance on intricate queries. Another technique, retrieval-augmented generation, provides relevant and helpful examples as a few-shot to guide SQL generation \cite{rag}. These approaches have proven effective on general Text-to-SQL tasks but they have not yet been studied rigorously on domain-specific Text-to-SQL problems. Retrieval Augmented Fine-tuning (RAFT) introduces a novel fine-tuning technique that improves the in-domain performance of RAG while integrating domain-specific knowledge \cite{raft}. 

Through our work, we delve into the application of these techniques for the EHRSQL-2024 task.

\section{Shared Task and Dataset}
The EHRSQL-2024 shared task \cite{ehrsql2024} is aimed at creating a reliable SQL for answering questions posed in natural language on the MIMIC-IV demo database. The MIMIC-IV database consists of anonymized electronic health records of patients admitted to the Beth Israel Deaconess Medical Center. These records primarily cover two modules: hospital records and ICU records. The publicly available demo version of the database contains a subset of patient records for 100 individuals. It consists of 17 tables from both modules, encompassing a total of 109 columns.

\subsection{Task Definition}

The task aims to accurately generate SQL queries for answerable questions and predict null $(\phi)$ for unanswerable ones. Each correct answer earns a score of 1, while incorrect answers receive a score of $-c$, where $c$ is the associated cost. The overall score $RS$ for a cost $c$ and prompt parameter $\theta$ can be expressed as below.

\begin{equation}
\begin{aligned}[b]
  \label{eq:rsc}
  RS_\theta(C) = \Sigma^{N}_{i=1} \mathbbm{1}({E(LLM_\theta(Q_i)) = E(GT_i)}) \\ - C * \mathbbm{1}({E(LLM_\theta(Q_i)) \neq E(GT_i)})
\end{aligned}
\end{equation}

where $LLM$ represents the model that generates SQL based on a given question $Q_i$. $GT_i$ denotes the ground truth SQL query for the question, and $E$ signifies the executed value of the SQL query when run on a specific database. $\mathbbm{1}$ is the indicator function.

The objective of this task is to find the optimal value of $\theta$ at a cost $c$ with respect to the function $RS_\theta(C = c)$.

\subsection{Dataset}
The dataset contains a combination of answerable and unanswerable questions across three subsets: train, valid, and test. Table \ref{tab:dataset_stats} provides an overview of the composition of each subset.

\begin{table}
  \centering
  \begin{tabular}{c c c}
    \hline
     & \textbf{Total Samples} & \textbf{\% Unanswerable} \\
    \hline
    \textbf{Train} & 5124 & 8.78 \% \\
    \textbf{Valid} & 1163 & 19.95 \% \\
    \textbf{Test} & 1167 & 19.97 \% \\
    \hline
  \end{tabular}
  \caption{EHRSQL-2024 Dataset Statistics}
  \label{tab:dataset_stats}
\end{table}

\section{Approach}
The reliable Text-to-SQL solution is decomposed into two stages as follows.

\subsection{SQL Generation}
To begin, we concentrate solely on boosting the number of accurately produced SQLs without being concerned with reducing the number of incorrect responses. As a result, the objective function becomes:

\begin{equation}
  \label{eq:rs0}
  RS_\theta(C=0) = \Sigma^{N}_{i=1} \mathbbm{1}({E(LLM_\theta(Q_i)) = E(GT_i)})
\end{equation}

Maximizing the success and minimizing hallucinations of the LLMs generation task require the provision of the correct context. To achieve this, the following information is essential regarding the task at hand:
\begin{itemize}
    \item \textbf{Database Schema} Comprising tables, columns, and their interrelationships, the database schema serves as a blueprint for the data stored in the database. This information guides the LLM in selecting the appropriate tables and columns.
    \item \textbf{Database Column Values} The actual values stored in the table columns offer additional information. This helps the LLM comprehend and perform operations such as data validation, manipulation, and filtering
    \item \textbf{Training Data} Providing questions (with corresponding SQL answers) similar to the current question aids the LLM in comprehending query formats, syntax, semantics, ambiguity resolution, and bridging the real-world knowledge gap with EHRSQL.
\end{itemize}

To produce SQL queries for each question, we employ an \textbf{in-context learning} approach. Here, the LLM is provided with similar question-SQL pairs, along with the relevant database content. To retrieve similar questions from the training data, we calculate cosine similarities between the evaluation question embedding and the training question embeddings.

We utilize the AnglE model based on BERT, which aims to minimize the angle difference in a complex space \cite{angle}. This approach helps overcome the negative impact caused by the saturation zone of the cosine function. The AnglE embedding model ranks among the top 10 in the Massive Text Embedding Benchmark (MTEB), encompassing eight embedding tasks and 58 datasets \citep{mteb}. While AnglE effectively captures the semantic similarity between the intent of questions, it faces challenges in capturing the similarity between clinical terminology, which is also crucial for this task.

To bridge this gap, we combine AnglE embeddings with PubMedBERT embeddings \cite{pubmedbert}, trained on the PubMed literature. This allows us to enhance the system’s ability to capture clinical terminology. Since embedding similarity scores are not directly comparable across different models due to varying dimensionality, we perform z-normalization to ensure comparability.
Algorithm \ref{alg:multi-embedding} provides an overview of how we retrieve the top N similar questions for a given question using two different embedding models.

\begin{algorithm*}
\caption{Multi-Embedding Retrieval}
\label{alg:multi-embedding}
    \KwData{train\_questions, test\_questions, N }
    \KwResult{Similar train questions for test questions}
    \tcp{Same size as test\_questions}
    \tcp{Each element contains top N similar train questions and scores}
    result $\gets$ []\;
    \ForEach{embed\_model $\in$ M}{
    train\_embeddings $\gets$ create\_embeddings(embed\_model, train\_questions)\;
    test\_embeddings $\gets$ create\_embeddings(embed\_model, test\_questions)\;
    questions, scores $\gets$ compute\_similarity(test\_embeddings, train\_embeddings, top\_n=N)\;
    $\mu \gets$ compute\_mean(scores)\;
    $\sigma \gets$ compute\_std(scores)\;
    z\_scores $\gets$ (scores $- \mu$) $/ \sigma$\;
    \tcp{Sort and merge top N current questions with result}
    \tcp{If questions overlap, update with max score}
    result $\gets$ sort\_and\_merge(result, questions, z\_scores)\;
    }
\end{algorithm*}

To generate the SQL, we employed ICL and fine-tuning approaches with a consistent prompt template. A shorter version of the final prompt template is provided below for reference. For ICL, we utilized pre-trained models, such as GPT-4 \cite{gpt4} and Claude-3 Opus \cite{opus}, with their default settings for temperature, top\_p, and top\_k parameters. To fine-tune GPT-3.5, we leveraged the retrieval augmented fine-tuning (RAFT) technique. For each training question, we generated similar questions using the multi-embedding retrieval approach while maintaining the prompt template. Given the limited size of the training set, we conducted fine-tuning with default parameters for only one epoch to prevent overfitting.
\\
\begin{mdframed}
\begin{center} {\bf Prompt Template} \end{center}
This is a task converting a natural language question to an SQLite query for a database. You will be provided with the schema of the SQLite database followed by a few examples. You need to generate the SQLite query for a given question and you may return  "null" if the question cannot be answered.
\begin{verbatim}
[Database Tables]
CREATE TABLE patients
(
  row_id int not null primary key, -- 42
  subject_id int not null unique, -- 201
  gender varchar(5) not null, -- 'm'
  dob timestamp(0) not null,
  dod timestamp(0)
);
...
...
[Examples]
[Q]  : How many patients are there in 
       total?
[SQL]: SELECT count(subject_id) FROM 
       patients
[Q]  : What is the gender of patient
       1002?
[SQL]: SELECT gender FROM patients 
       WHERE subject_id = 1002
[Q]  : What is the date of birth of 
       patient 1002?
SQL: 
\end{verbatim}
\end{mdframed}

\

\begin{figure*}[!ht]
\centering
\caption{SQL Generation Process}
\includegraphics[width=16cm]{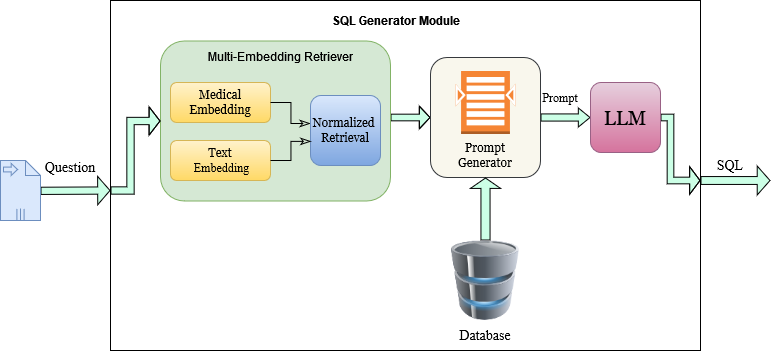}
\label{fig:generation}
\end{figure*}

Figure \ref{fig:generation} illustrates the complete process of generating SQL using an LLM post training for a given question.

\subsection{SQL Validation}
LLMs have a tendency to generate inaccurate and imaginary responses, regardless of the quality of the context they are provided. Therefore, we implement a second stage using an ensemble approach to eliminate errors generated during the initial generation stage. To verify whether the SQL generated by a two-model or three-model ensemble is correct, each query result is obtained by evaluating it against the database. Subsequently, the results are compared, and a match among all the results qualifies the query as correct.

\section{Results}
In this section, we present the comparison of the reliability scores of the individual models followed by ensembles.

\subsection{Individual Models}
Table \ref{tab:first_stage} presents the reliability scores along with the percentage of unanswered questions for each model i.e. GPT-4, Claude-3 Opus and fine-tuned GPT-3.5.

\begin{table}
  \centering
  \begin{tabular}{c c c c}
    \hline
    \textbf{Model} & \textbf{RS0} & \textbf{RS10} & \textbf{Unanswered \%} \\
    \hline
    GPT-4 & 88.51 & 40.53 & 25.71  \\
    FT GPT-3.5 & 88.08 & 22.96 & 23.14 \\
    Opus & 88.94 & 18.68 & 22.28 \\
    \hline
  \end{tabular}
  \caption{Reliability Scores of Individual Models}
  \label{tab:first_stage}
\end{table}

Overall, Claude-3 Opus answered the most number of questions correctly while also answering them wrong more than others which led to the lowest RS10. GPT-4 appears to be more conservative in generating SQLs and has generated the most nulls. As refraining from generating for unanswerable questions is more important in this task, this led to achieving the best score on RS10 for GPT-4. Although the GPT-3.5 model is significantly less performant than GPT-4, the fine-tuned version brought the generation capability close to the GPT-4 model.

\subsection{Ensemble}
To select the ensemble model that achieves the best performance, we comprehensively evaluated all possible combinations of 2-model and 3-model ensembles. Table \ref{tab:ensemble} provides a detailed comparison of the reliability scores achieved by these various model ensembles.

\begin{figure}[H]
\caption{Individual vs Ensemble Models}
\includegraphics[width=8cm]{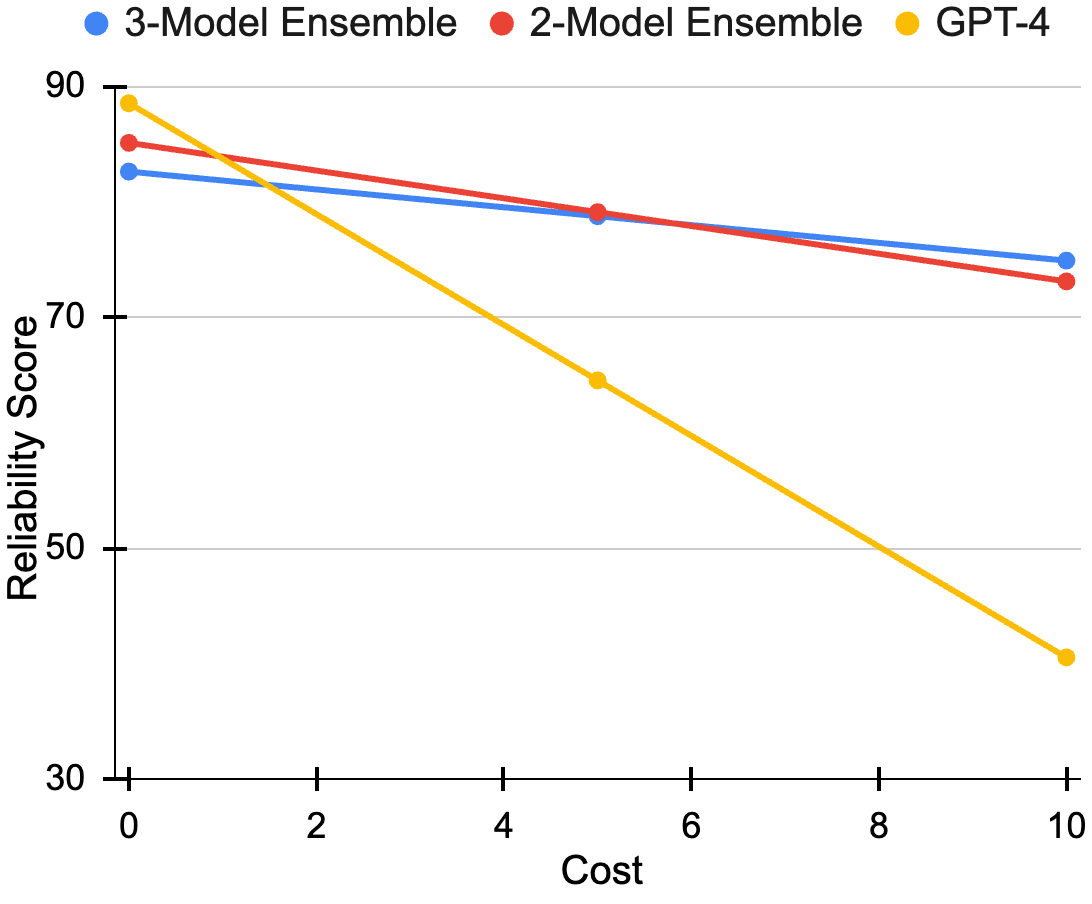}
\label{fig:comparison}
\end{figure}

Among the 2-model ensembles, the combination of fine-tuned GPT-3.5 and Claude-3 Opus achieved the highest RS10 score, outperforming other models. Notably, the ensemble approach involving the fine-tuned GPT-3.5 model exhibited a significant reduction in errors compared to pre-trained models. This finding suggests that the fine-tuned model produces distinct errors from the pre-trained models, thus maximizing the validation benefits of ensemble approaches. The 3-model ensemble, however, achieved the best RS10 score among all approaches. To illustrate the effectiveness of Ensemble models, Figure \ref{fig:comparison} demonstrates the reliability scores of top-performing models from the individual, 2-model ensemble, and 3-model ensemble categories. When comparing against a stand-alone model, both 2-model and 3-model ensembles substantially minimize errors and obtain roughly equivalent but large RS10 scores. These results clearly demonstrate that ensemble approaches are effective validation mechanisms for creating reliable and accurate SQL generation systems.

\begin{table}
  \centering
  \begin{tabular}{c c c c}
    \hline
    \textbf{Ensemble} & \textbf{RS0} & \textbf{RS10} \\
    \hline
    GPT-4 + Opus & 84.57 & 65.72  \\
    FT GPT-3.5 + GPT-4 & 84.83 & 71.97 \\
    FT GPT-3.5 + Opus & 85.08 & 73.09 \\
    All & 82.6 & 74.89 \\
    \hline
  \end{tabular}
  \caption{Reliability Scores of Ensemble Models}
  \label{tab:ensemble}
\end{table}

\section{Ablation Study}
To assess the significance of each parameter in the final prompt employed for ICL and fine-tuning, we conduct an ablation study. In these experiments, we focus solely on pre-trained models because fine-tuning experiments are more expensive and time-consuming. To accelerate the process and maintain costs, we leverage GPT-3.5, a compact and less potent yet faster variant of the GPT family.
Through these experiments, we extrapolate the efficacy of each parameter for prompting using more robust and advanced models such as GPT-4 and Claude-3 Opus. Table \ref{tab:ablation} provides the reliability scores obtained by progressively constructing a prompt with varying levels of complexity.

\begin{table}
  \centering
  \resizebox{\columnwidth}{!}{%
  \begin{tabular}{c c c c}
    \hline
    \textbf{Prompt Type} & \textbf{\makecell{Executable\\ \%}} & \textbf{RS0} & \textbf{RS10} \\
    \hline
    \makecell{No Few-shot} & 83.84 & 32.84 & -440.06 \\
    \hline
    \makecell{One Embedding\\ Few-Shot} & 95.89 & 66.98 & 7.65 \\
    \hline
    \makecell{Two Embeddings\\ Few-Shot} & 98.34 & 69.13 & 11.52 \\
    \hline
    \makecell{Two Embeddings Few-\\Shot + Column Values} & 95.71 & 69.3 & 15.99 \\
    \hline
  \end{tabular}%
  }
  \caption{Reliability Scores of GPT-3.5 with Different Prompt}
  \label{tab:ablation}
\end{table}

Incorporating few-shot examples in the prompt has substantially improved both the executable queries and reliability scores. This demonstrates the critical role of ICL with few-shot in Text-to-SQL tasks, particularly in the context of EHRSQL. The one-embedding few-shot experiment employs non-medical AnglE embeddings \cite{angle}, while the two-embeddings few-shot additionally leverages PubMedBERT \cite{pubmedbert}. It is evident that adding medical embeddings enhances all metrics by a good margin. While adding column values to the few-shot prompt decreased executable queries potentially leading to an increase in RS10, it also showed an improvement in RS0, indicating its usefulness as a signal. Through these experiments, we arrived at the final prompt, which enabled us to develop a highly reliable Text-to-SQL system.

\section{Conclusion}
Our work primarily aims to enhance the reliability of SQL generation, which is of paramount importance for the EHRSQL-2024 shared task. Although in-context learning with advanced LLMs such as GPT-4, Claude-3 Opus, or fine-tuning GPT-3.5 yields excellent RS0, errors still seem inevitable. The model's ability to solve a specific task is heavily influenced by the training data. Repeatedly generating using the same prompt (or) the same model to validate often fails to minimize errors since hallucinations mainly originate from the training data. Fine-tuning GPT-3.5 resulted in different error tendencies compared to pre-trained models, even when using the same prompt. Therefore, ensemble LLMs, particularly those with a fine-tuned model, offer a superior approach for SQL validation, improving robustness and reliability. This approach has also secured us 2nd place in the competition.

\section{Limitations and Risks}
Our approach, while successful in this context, requires careful planning for real-world deployment due to certain limitations. Fine-tuning GPT-3.5 is computationally expensive and necessitates high-quality training data.  Ensemble methods, though powerful for validation, introduce trade-offs in terms of cost and complexity.  Crucially, it's vital to evaluate potential biases inherited from the LLM's training data to ensure fair and reliable performance in practical applications.

\bibliography{custom}

\end{document}